\documentclass{elsart}
\usepackage{graphicx}
\usepackage{amssymb}

\def\Journal#1#2#3#4{{#1} {#2} (#4) #3}

\def\numpb{1,520}
\def\numUM{8}
\def\numUL{1,512}

\begin{document}

\begin{frontmatter}
  \title{
    Measurement of the cosmic-ray low-energy antiproton spectrum
    with the first BESS-Polar Antarctic flight
  }

\author[Kobe]{K. Abe\thanksref{icrr}},
\author[ISAS]{H. Fuke},
\author[KEK]{S. Haino\thanksref{infn}\corauthref{cor}},
\corauth[cor]{Corresponding author.}
\ead{haino@post.kek.jp}
\author[NASA]{T. Hams\thanksref{usra}},
\author[Kobe]{A. Itazaki},
\author[UM]{K. C. Kim},
\author[KEK]{T. Kumazawa},
\author[UM]{M. H. Lee},
\author[KEK]{Y. Makida},
\author[KEK]{S. Matsuda},
\author[KEK]{K. Matsumoto},
\author[NASA]{J. W. Mitchell},
\author[NASA]{A. A. Moiseev},
\author[UM]{Z. Myers\thanksref{tech}},
\author[Tokyo]{J. Nishimura},
\author[KEK]{M. Nozaki},
\author[Kobe]{R. Orito\thanksref{mpi}},
\author[Denver]{J. F. Ormes},
\author[NASA]{M. Sasaki\thanksref{usra}},
\author[UM]{E. S. Seo},
\author[Kobe]{Y. Shikaze\thanksref{jaea}},
\author[NASA]{R. E. Streitmatter},
\author[KEK]{J. Suzuki},
\author[Kobe]{Y. Takasugi},
\author[Kobe]{K. Takeuchi},
\author[KEK]{K. Tanaka},
\author[ISAS]{T. Yamagami},
\author[KEK]{A. Yamamoto},
\author[ISAS]{T. Yoshida},
\author[KEK]{K. Yoshimura}

\address[Kobe]{Kobe University, Kobe, Hyogo 657-8501, Japan}
\address[ISAS]{Institute of Space and Astronautical Science,
 Japan Aerospace Exploration Agency (ISAS/JAXA),
 Sagamihara, Kanagawa 229-8510, Japan}
\address[KEK]{High Energy Accelerator Research Organization (KEK),
 Tsukuba, Ibaraki 305-0801, Japan}
\address[NASA]{National Aeronautics and Space Administration,
 Goddard Space Flight Center (NASA/GSFC), Greenbelt, MD 20771, USA}
\address[UM]{IPST, University of Maryland, College Park, MD 20742, USA}
\address[Tokyo]{The University of Tokyo, Bunkyo, Tokyo 113-0033 Japan}
\address[Denver]{University of Denver, Denver, CO 80208, USA}

\thanks[icrr]{Present address: Kamioka Observatory, ICRR, 
              The University of Tokyo, Kamioka, Gifu 506-1205, Japan}
\thanks[infn]{Present address: Istituto Nazionale di Fisica Nucleare (INFN), 
              Perugia 06123, Italy}
\thanks[usra]{Also at: CRESST/USRA, Columbia, MD 21044, USA}
\thanks[tech]{Present address: Physics Department, Technion - Israel 
  Institute of Technology, Technion City, Haifa 32000, Israel.}
\thanks[mpi]{Present address:  
  Max-Planck-Institut f\"{u}r Physik, M\"{u}nchen 80805, Germany}
\thanks[jaea]{Present address: Japan Atomic Energy Agency (JAEA),
  Tokai-mura, Naka-gun, Ibaraki 319-1195, Japan}

\begin{abstract}
The BESS-Polar spectrometer had its first successful balloon flight 
over Antarctica in December 2004. During the 8.5-day long-duration flight, 
almost 0.9 billion events were recorded and \numpb\ antiprotons were 
detected in the energy range 0.1--4.2~GeV. 
In this paper, we report the antiproton spectrum obtained, 
discuss the origin of cosmic-ray antiprotons, 
and use antiproton data to probe the effect of charge-sign-dependent 
drift in the solar modulation. 
\end{abstract}

\end{frontmatter}

\section{Introduction}
\label{sec:introduction}

Antiproton spectra have been measured by BESS 
(Balloon-borne Experiment with a Superconducting Spectrometer) 
in a series of flights from Lynn Lake, 
Canada providing reasonable statistics above 1 GeV~\cite{Yoshimura95,
  Moiseev97,Matsunaga98,Orito00,Maeno01,Asaoka02,Haino05}. Below 1 
GeV statistics are limited and the effects of solar modulation are greater. 
The spectrum has a distinct peak around 2 GeV, 
showing the characteristic feature of secondary antiprotons 
produced by the interaction of Galactic cosmic-rays 
with the interstellar medium. 
The energy spectrum of these secondary antiprotons should 
decrease rapidly toward lower energies 
reflecting the kinematic constraints on antiproton 
production~\cite{Gaisser92,MitsuiD96,Moskalenko02} 
and toward higher energies reflecting the steep power-law spectra 
of primary particles producing the antiprotons.
In addition to these secondary antiprotons, there might be a source 
of primary antiprotons. Such sources have been suggested to result from the 
evaporation of primordial black holes (PBH) or from the annihilation of 
neutralino dark matter~\cite{Hawking74,Maki96,Mitsui96,Bergstroem99}. 
The spectrum of antiprotons from a primary source might have a peak 
in the energy region below 1~GeV, giving a flatter composite 
spectrum with excess flux compared to the purely secondary 
spectrum~\cite{Moskalenko02,Mitsui96}.
The influence of a low energy primary peak would be most evident at 
solar minimum~\cite{Maki96,Mitsui96} and BESS (1995+1997) measurements 
during this period~\cite{Matsunaga98,Orito00} suggested this possibility. 
On the other hand, BESS-1998~\cite{Maeno01} and subsequent 
measurements~\cite{Asaoka02,Haino05}, which were taken after the solar 
minimum period, are more consistent with pure secondary nature. 
In order to resolve the important questions regarding possible 
novel sources, we undertook the higher precision 
measurements reported here. 

In addition, one can use the charge-sign difference of antiprotons to 
explore effects of drift in the solar modulation. 
A detailed understanding of the effects of solar modulation 
is important to establish the existence of any primary sources. 
During the positive polarity phase of solar activity before 2000
the measured antiproton-to-proton (\={p}/p) ratio~\cite{Maeno01} 
showed no distinctive variation. After the reversal of the solar magnetic 
field in 2000, a sudden increase of the \={p}/p ratio~\cite{Asaoka02} 
was clearly observed. 
Our measurements, as well as measurements of the positron 
fraction (e$^+$/(e$^+$+e$^-$))~\cite{Clem04}, 
generally support recent calculations~\cite{Moskalenko02,Bieber99} 
incorporating steady-state drift models and charge-dependent effects 
of solar modulation. 
Protons and antiprotons have significantly different interstellar 
spectra and the drift directions are opposite because of the opposite 
charge sign. The combination of these effects implies 
that the \={p}/p ratio should display a more interesting 
evolution~\cite{Bieber99} during 2000-2010 than it did during the 1990's. 
The secondary antiprotons are also useful to probe the average cosmic-ray 
proton spectrum over a large region of the Galaxy~\cite{Strong04}.

The BESS-Polar experiment was proposed as an advanced BESS program 
of long-duration balloon flights over Antarctica 
and has been prepared since 2001~\cite{Yamamoto02A,Mitchell04,Yoshida04} 
to further investigate elementary particle 
phenomena in the early Universe through a precise measurement of 
the low-energy antiproton spectrum and to search for primary antinuclei 
in cosmic rays. 
We report here a measurement of the cosmic-ray antiproton spectrum 
based on \numpb\ antiproton events detected during an 8.5-day 
flight of the BESS-Polar spectrometer performed in December 2004. 
We discuss
the origin of cosmic-ray antiprotons, and 
the solar modulation effect based on the measured \={p}/p ratio.

\section{Spectrometer}
\label{sec:spectrometer}

\begin{figure}
  \begin{center}
    \includegraphics[width=8cm]{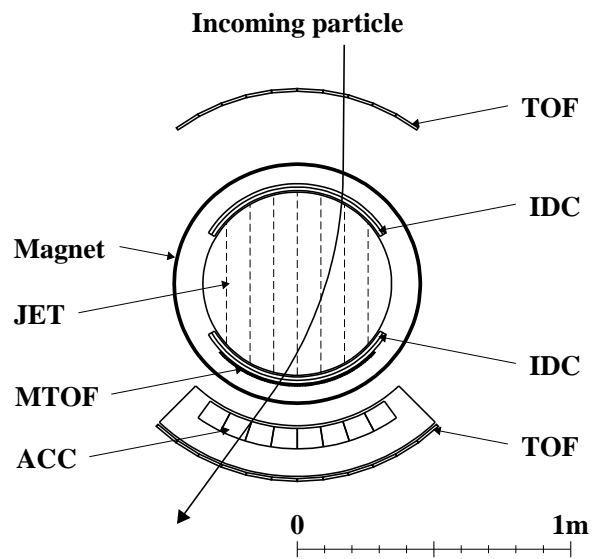}
    \caption
    {Cross-sectional view of the BESS-Polar spectrometer.}
    \label{fig:spectrometer}
  \end{center}
\end{figure}

The BESS-Polar superconducting spectrometer, 
shown in Fig.~\ref{fig:spectrometer}, has been developed 
to reduce material thickness along the particle trajectory and 
to meet the severe requirements for long duration balloon flights 
over Antarctica~\cite{Yamamoto02A,Yoshida04}. 
The basic instrument concepts, such as the cylindrical configuration 
with an open and wide acceptance and redundant measurements for particle 
identification, are inherited from the BESS spectrometer~\cite{Ajima00}. 
A very low instrumental energy cutoff for antiprotons was 
achieved with a new thin-walled (2.46~g/cm$^2$/wall including cryostat) 
superconducting magnet~\cite{Yamamoto02B,Makida06} using a new high-strength 
aluminum-stabilized superconductor. 
In addition, the outer pressure vessel was eliminated and the detectors 
were reconfigured. Low-energy particles only have to 
traverse 4.5~g/cm$^2$ of instrument material to be detected, about one 
quarter of that in the previous BESS spectrometer~\cite{Ajima00,Haino04A}. 
With these changes, 
the lowest energy for antiproton detection has been reduced to 
0.1~GeV at the top of atmosphere (TOA). 

A uniform field of 0.8 Tesla is produced by the superconducting coil, 
and its continuous operation is maintained for over 11 days with a 400-liter 
liquid helium reservoir. 
A JET-cell type drift chamber (JET) and two inner drift chambers (IDCs), 
which were also used in the BESS-TeV spectrometer~\cite{Haino04A,Haino04B}, 
are placed inside the warm bore (0.80~m in diameter and 1.4~m in length).
The JET and IDCs are filled with CO$_2$ gas, and fresh gas is circulated 
using a semi-active flow control system. 
Tracking of an incident particle in the $x-y$ plane (perpendicular to the 
magnetic field) inside the JET and IDCs is performed by fitting 
up to 52 hit-points, each with 150 $\mu$m resolution, 
resulting in a magnetic-rigidity  
($R\equiv p$c$/Ze$, momentum divided by electric charge) 
resolution of 0.4~\% at 1 GV, and a maximum detectable rigidity 
(MDR) of 240 GV. 
Tracking in the $z$ coordinate (parallel to the magnetic field direction) 
is done by fitting points inside IDCs measured by vernier pads 
with an accuracy of 1.0~mm and points in JET measured 
by charge division with an accuracy of 37~mm. 
The continuous and redundant 3-dimensional tracking enables BESS-Polar 
to distinguish backgrounds such as tracks having interaction or scattering. 
A truncated mean of the integrated charges of the hit-pulses from the 
JET also provides an energy loss (d$E$/d$x$) measurement 
with a resolution of 10~\%. 

The top and bottom scintillator hodoscopes, the time-of-flight (TOF) 
detectors, are used to determine the incident particle velocity, 
$\beta = v/{\rm c}$. They also make two independent d$E$/d$x$ measurements. 
The scintillator hodoscopes consist of 10 top and 12 bottom plastic 
scintillators with a cross section of 96.5~mm (width) $\times$ 
10~mm (thickness) and with photo-multiplier tubes (PMTs) attached by acrylic 
light guides at each end. The timing resolution of each hodoscope is 110 ps, 
resulting in a $\beta^{-1}$ resolution of 3.3~\%.
In addition, a thin scintillator middle-TOF (MTOF) is 
installed on the lower surface of the solenoid bore to detect low energy 
particles which cannot penetrate the magnet wall. 
The MTOF consists of 64 plastic scintillator bars with a cross section of 
10~mm (width) $\times$ 5~mm (thickness) read by multi-anode PMTs. 
The $\beta$ of the low energy antiprotons can be measured as 
a combination of top-middle TOF with a $\beta^{-1}$ resolution 
of 4.5~\% below 0.6~GV (0.2~GeV for proton and antiproton). 
A \v{C}erenkov counter with silica-aerogel radiator (ACC) 
is installed below the magnet. 
The radiator is selected to have a refractive index of 1.02 
in order to veto $e^{-}/\mu^{-}$ backgrounds up to 4.2 GeV. 
The top and bottom TOF hodoscopes and the ACC operate at ambient pressure.

The event data acquisition sequence is initiated by a coincidence of 
signals in the top-bottom TOF or top-middle TOF hodoscopes~\cite{Sasaki05}. 
Based on digitized detector information sent from the front-end electronics, 
the event data are built and recorded to onboard hard disk drives 
with a total capacity of 3.5 Tera Bytes. 
During the Antarctic flight, the data acquisition rate was around 1.4~kHz, 
and the dead time was 150 $\mu$s per event (20~\% of the data-taking time). 
To supply electric power to the electronics onboard the payload, 
a solar cell array of 900~W capacity was mounted on 
an omni-directional octagonal frame around the payload. 
The total power consumption of the spectrometer was 420~W. 

\section{Data Analysis}
\label{sec:analysis}

\subsection{Balloon flight observation}
\label{sub:observation}
The first BESS-Polar long duration balloon flight over Antarctica 
was launched from Williams Field 
($77^{\circ}51.8'$S, $167^{\circ}5.4'$E) near McMurdo Station on 
December 13, 2004~\cite{Yoshida05}. 
During the flight, some PMTs of the TOF hodoscopes had to be turned off 
because they were drawing excessive current, and the usable geometrical 
acceptance was consequently reduced to 73~\% of the design value 
(0.157~m$^2$sr at 0.2~GeV and 0.166~m$^2$sr at 2.0~GeV). 
The flight was terminated on December 21 and the payload landed at the south 
end of the Ross Ice Shelf ($83^{\circ}6.0'$S, $155^{\circ}35.4'$W) 
after a continuous observation period of 8.5~days. 
The flight trajectory was close enough to the South magnetic pole 
that the geomagnetic cutoff rigidity was below 0.2~GV, lower 
than the lowest detection limit of the spectrometer. 
During the live data-taking time of 507,075 seconds 
at an average floating altitude of 38.5~km 
(residual atmosphere of 4.3~g/cm$^2$), 
894,482,590 cosmic-ray events were accumulated 
without any online event selection as 
2.14 terabytes of data recorded on the hard disk drives. 

\subsection{Event selection}
\label{sub:selection}
Since the spectrometer is cylindrically symmetric, it may be assumed 
that antiprotons behave exactly like protons in the instrument 
except for the sign of their deflection in the magnetic field 
and their inelastic interactions. 
Thus, all the selection criteria were defined 
based on the measured properties of protons. 
Initially, events were selected with (1) a single track and a downward-going 
particle fully contained in the fiducial region of the tracking volume, 
(2) only one or two hits each in the top and bottom TOF hodoscopes, 
(3) the hit position at the TOF hodoscopes consistent with the 
extrapolated track inside the JET and IDCs, 
(4) cosine of zenith angle of the incident particle 
larger than 0.8, and 
(5) either top or bottom TOF hodoscopes read by PMTs at both ends. 
As a consequence of (5), the effective geometrical acceptance 
had to be further reduced to be 40~\% of the design value. 

The selection efficiency was estimated using a Monte Carlo (MC) simulation 
by applying the same selection criteria to the simulated events 
as to the observed data. 
The MC simulation was based on a GEANT3/GHEISHA code~\cite{Geant3,Gheisha} 
and tuned to reproduce the results of an accelerator beam test of 
the previous BESS spectrometer~\cite{Asaoka01} in which the detector 
configuration and materials are similar to the BESS-Polar spectrometer. 
The efficiency varied from 
84.2$\pm$5.0~\% at 0.2~GeV to 90.1$\pm$5.0~\% at 2.0~GeV. 
The systematic error of the efficiency was determined using the  
accelerator beam test data. 
Each analog-to-digital board for TOF has an individual dead time due to 
the period of switching reference capacitors used for baseline subtraction. 
Events digitized during the dead time have incorrect charge data 
and were rejected to ensure the quality of the data. The efficiency of 
surviving this cut was estimated as 77.8$\pm$0.1~\% at 0.2~GeV 
and 78.8$\pm$0.1~\% at 2.0~GeV. 
Here and in the selections described below, a sample of the proton data 
has been used to determine the efficiencies. 

The following track quality cuts were applied: 
(1) the reduced $\chi^{2}$ in the $x$--$y$ and $y$--$z$ track fitting 
to be less than 5, 
(2) the fitting error on the curvature (inverse rigidity) 
to be less than 0.015~GV$^{-1}$, 
(3) the track fitting path length to be longer than 500~mm, and 
(4) the residual between hit position at TOF obtained from the time difference 
in two PMTs and the extrapolated track of the JET to be less than 50~mm. 
The quality cut efficiency was estimated as 94.7$\pm$0.2~\% at 0.2~GeV 
and 91.7$\pm$0.1~\% at 2.0~GeV.

\subsection{Particle identification}
\label{sub:pid}
\begin{figure}
  \begin{center}
    \includegraphics[width=12cm]{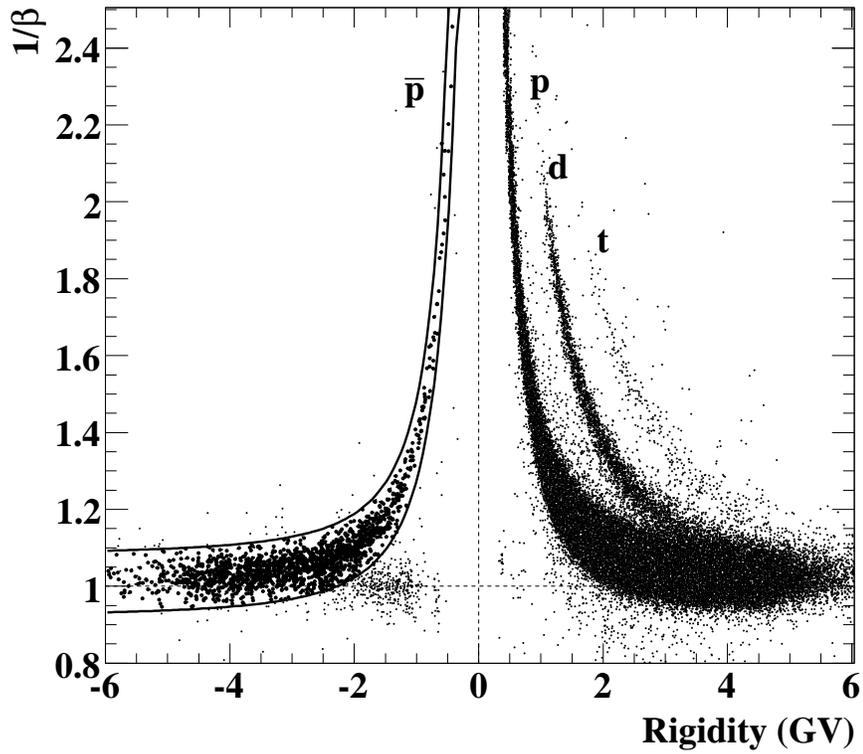}
    \caption
    {The $\beta^{-1}$ versus rigidity plot, and antiproton selection band. 
     The same band but opposite rigidity sign is applied to select protons.
     For the negative rigidity, all the events after \v{C}erenkov veto cuts 
     and JET d$E$/d$x$ cut are shown. For the positive rigidity, 
     0.5~\% of the events after \v{C}erenkov veto cuts are shown.
     Antiproton candidates are shown with dots of larger size.}
    \label{fig:idplot}
  \end{center}
\end{figure}

Proton and antiproton candidates were identified 
with a combination of a charge selection 
and a mass selection as follows:
(1) Particle charge, $Z$ is identified by d$E$/d$x$ measurements. 
The d$E$/d$x$ measurements with both top and bottom TOF hodoscopes  
were required to be inside a band defined as a function of rigidity 
corresponding to singly-charged particles. 
(2) Particle mass $m$ is reconstructed with rigidity $R$, velocity $\beta$ 
and $Z$ as $m=ZeR\sqrt{1/\beta^2-1}$. 
It was required that $1/\beta$ be inside 
a band defined as a function of rigidity as shown in Fig.~\ref{fig:idplot}, 
so that the reconstructed mass was consistent with 
that of a proton or antiproton. 
The d$E$/d$x$ and $\beta$ bands for antiprotons were defined 
in the same way as for protons except for the rigidity sign. 
The selection efficiency of each d$E$/d$x$ band cut was estimated 
as a fraction of the number of selected events among a proton sample selected 
by the other independent d$E$/d$x$ selections. The net efficiency 
for the two d$E$/d$x$ band cuts was 96.5$\pm$0.1~\% at 0.2~GeV 
and 96.4$\pm$0.1~\% at 2.0~GeV. 
The selection efficiency of the $1/\beta$ band cut was estimated as 
99.5$\pm$0.2~\% at 0.2~GeV and 97.0$\pm$0.1~\% at 2.0~GeV. 

In order to eliminate e$^-$ and $\mu^-$ backgrounds that mimic 
relativistic antiproton candidates with $\beta>0.9$, 
\v{C}erenkov veto cuts were applied to select  
(3) the particle trajectory to be inside an aerogel fiducial volume, and 
(4) the \v{C}erenkov outputs to be less than a threshold. 
The rejection factor was estimated as (8.9$\pm$0.2)$\times$10$^2$ using 
a relativistic proton sample with rigidity larger than 20~GV. 
The efficiency including a loss of fiducial volume by 18~\% 
was estimated as 75.8$\pm$0.1~\% at 0.4~GeV 
and 62.3$\pm$0.1~\% at 2.0~GeV. 

Because of the relativistic rise of d$E$/d$x$ in the JET gas, 
ultra-relativistic particles ($\beta\rightarrow$1) have d$E$/d$x$ about 1.4 
times higher than minimum ionizing particles. A tight cut on d$E$/d$x$ 
measured with the JET can eliminate part of the e$^-$ and $\mu^-$ backgrounds. 
(5) The d$E$/d$x$ measurements with the JET were required 
to be inside a band defined as a function of rigidity. 
The rejection factor was estimated to be 1.75$\pm$0.03 at 2.0~GeV 
and 4.85$\pm$0.08 at 4.0~GeV, as a fraction of the rejected 
events among the e$^-$ and $\mu^-$ sample 
which was obtained with the same selection criteria as antiprotons excluding 
\v{C}erenkov veto cuts.
The efficiency was estimated to be 98.2$\pm$0.1~\% at 2.0~GeV 
and 96.9$\pm$0.1~\% at 4.0~GeV. 

Following the particle identification procedure, 
\numUL\ antiproton candidates were identified above 0.2~GeV as shown in 
Fig.~\ref{fig:idplot}.
Antiproton candidates that have wrongly reconstructed $\beta$ and 
lie above the selection band were rejected by the $1/\beta$ band cuts.  
The antiproton flux was obtained by correcting for the cut 
efficiency estimated using the proton sample. 
The event sample obtained with the antiproton selection criteria excluding 
\v{C}erenkov veto cuts and the JET d$E$/d$x$ cut 
consists of mostly ($>99.8$~\%) e$^-$ and $\mu^-$ background events. 
The number of such events divided by the 
total rejection factors defined above was used as an estimation of the 
number of contaminating backgrounds within the antiproton selection band.
The estimated number of background events (and the ratio to antiproton 
candidates) was 18.7$\pm$0.3 (8.5$\pm$0.1~\%) in the 1.7--2.1~GeV band 
and 8.1$\pm$0.1 (4.0$\pm$0.1~\%) in the 3.4--4.2~GeV band. 
In the region -1.5~GV~$<R<0$ and $0.9<1/\beta<1.1$, the remaining 
background events can be selected with only the $1/\beta$ band cut 
and the number of such events was 235, which was consistent with the
estimated number, 243$\pm$11.

Albedo and spillover from positive rigidity particles 
were negligible because of the high $\beta^{-1}$ 
and rigidity resolutions. 
To check against the ``re-entrant albedo'' 
background, we confirmed that the trajectories of all antiprotons 
could be traced numerically through the Earth's geomagnetic field back 
to the outside of the geomagnetic field~\cite{Honda04,igrf05}. 

\subsection{Selection with top-middle TOF below 0.2~GeV}
\label{sub:umconfig}
Some of the antiprotons below 0.2~GeV cannot reach the bottom TOF 
hodoscope due to energy and annihilation losses. 
To analyze such low energy events, we defined another set of selection 
criteria with a combination of the top and middle TOF hodoscopes. 
They are basically the same as those used with the top and bottom TOF, 
but the \v{C}erenkov veto cuts were not applied since at low energies 
antiproton candidates are well separated from e$^-$ and $\mu^-$ backgrounds 
using only the $1/\beta$ band cut. Using the top-middle TOF selection 
criteria, \numUM\ antiproton candidates were identified 
in the energy range 0.1--0.2~GeV. 

Stopping antiprotons can generate secondaries in the upward 
direction that make additional tracks inside the JET. Such events 
are rejected by the single-track selection. To recover such events, 
a search was made for antiproton candidates among multi-track events. 
The major backgrounds for this search are upward moving proton secondaries 
generated in the lower half of the spectrometer from interactions of high 
energy protons. 
Upward moving protons can usually be distinguished with the measured $\beta$, 
but if the incident proton and the upward moving secondary hit the same 
top TOF counter, the $\beta$ is not correctly measured. 
Such events were rejected by applying the following cuts:
(1) Stopping antiprotons cannot generate high energy secondaries 
because of the kinematical restriction. Hence for this search rigidities 
of all the tracks inside the JET were required to be less than 1~GV. 
(2) It was required that the distance between the two tracks at the top TOF 
to be larger than 150~mm. 
The efficiency of these cuts was estimated as 72.1$\pm$2.1~\% by applying 
the same cuts to MC generated antiproton events. 
A total of 31 multi-track antiproton candidates were found, 
but all of them were rejected by the cuts. 
According to the MC study, the total antiproton detection efficiency 
was increased by 11.1~\% at 0.1~GeV by accepting multi-track events. 
This increase was treated as a systematic error, but 
it was smaller than the statistical error (+41~\%/--34~\%)
of single-track antiproton candidates below 0.2~GeV. 

\subsection{Flux determination}
\label{sub:flux}
After the antiproton candidates were identified, energy-dependent 
corrections were applied for backgrounds and detection efficiency.
Then the absolute flux at the top of the instrument (TOI) 
was obtained by taking account of energy loss inside the spectrometer, 
live time, and geometrical acceptance. 
The energy of each particle at TOI was calculated 
by summing up the ionization energy losses inside the instrument as 
determined by tracing back the event trajectory. 
The effective geometrical acceptance 
was estimated using the MC code 
GEANT3~\cite{Geant3} and the simulation technique~\cite{Sullivan71} 
to be 0.114$\pm$0.001~m$^2$sr at 0.2~GeV and 0.121$\pm$0.001~m$^2$sr 
at 2.0~GeV. 
The error arising from uncertainty in the detector alignment 
was estimated to be 1~\%.

In order to obtain the flux at the top of the atmosphere (TOA), we applied 
a correction for survival probability of the flux reaching TOI from TOA, 
and a subtraction of the secondary component produced within the 
overlying atmosphere. The survival probability was estimated as 
89.1$\pm$2.0~\% at 0.2~GeV and 92.4$\pm$2.0~\% at 2.0~GeV 
based on total interaction lengths
 of 32.7$\pm$0.7~g/cm$^2$ at 0.2~GeV and 59.9$\pm$1.2~g/cm$^2$ at 
2.0~GeV~\cite{Stephens97,Stephens05}. 

The atmospheric secondary antiproton flux was estimated by solving 
simultaneous transport equations~\cite{Stephens97,Papini96}.
At low energies below 1~GeV, there exists a significant contribution of 
non-annihilating inelastic interactions or so-called ``tertiary'' antiprotons. 
The interaction length used was based on Stephen's model~\cite{Stephens05}, 
and the energy distribution of the tertiary production was tuned to reproduce 
the atmospheric antiproton flux measurements~\cite{Yamato06,Sanuki03}. 
The amount of secondary subtraction was 11.8$\pm$1.7~\% at 0.2~GeV and 
28.6$\pm$4.1~\% at 2.0~GeV. The relative error of 14.3~\% is composed 
of uncertainty in the residual air depth (5.0~\%), in 
the cross section of primary cosmic rays with air nuclei (8.9~\%), and 
in the tertiary production (10.0~\%). 

\section{Results and Discussions}
\label{sec:results}

\begin{figure}
  \begin{center}
    \includegraphics[width=12cm]{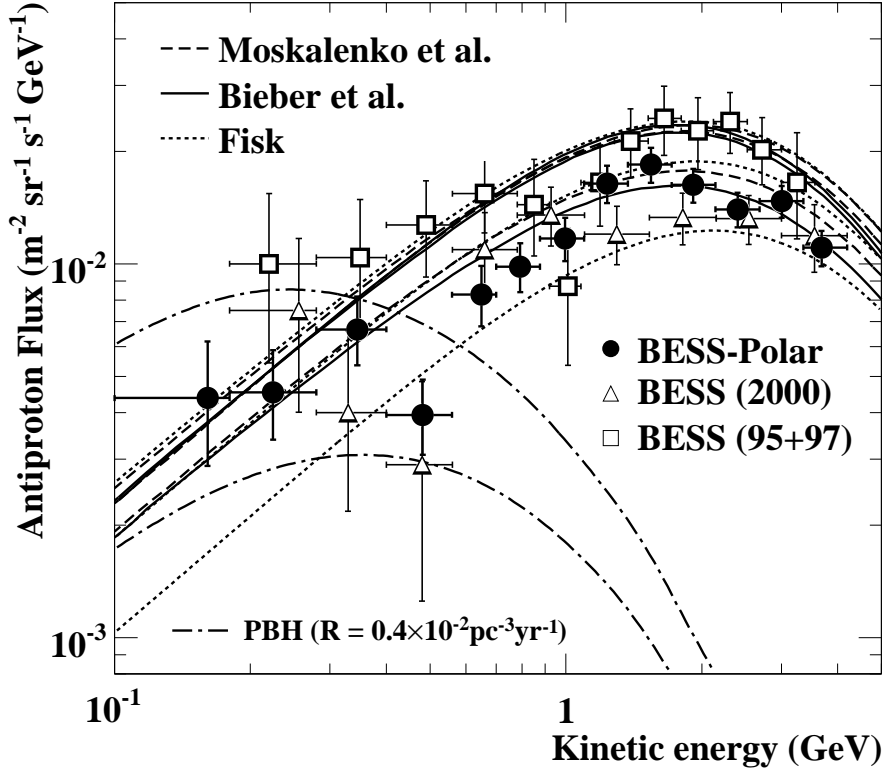}
    \caption
    {Antiproton flux at the top of the atmosphere obtained with the first 
     BESS-Polar flight together with results from previous BESS flights 
     around solar minimum (95+97)~\cite{Matsunaga98,Orito00}
     and maximum (2000)~\cite{Asaoka02}. 
     The solid curves are calculations of secondary antiproton spectra 
     with the Standard Leaky Box (SLB) model modulated with a steady state 
     drift model~\cite{Bieber99} by solar tilt angles and magnetic polarities 
     of (from top to bottom, the first two are very close) 
     10$^\circ(+)$, 10$^\circ(-)$, and 70$^\circ(-)$. The dashed curves 
     are calculations with the Diffusion plus Convection (DC) 
     model~\cite{Moskalenko02} modulated by (from top to bottom, the first 
     two are very close) 10$^\circ(+)$, 30$^\circ(-)$, and 70$^\circ(-)$.
     The dotted curves are calculations with the DC model~\cite{Moskalenko02} 
     modulated with a spherically symmetric model~\cite{Fisk71} 
     by (from top to bottom) 550~MV, 850~MV, and 1400~MV. 
     The dash-dot curves are calculations of antiproton spectra 
     from evaporation of primordial black holes with an explosion 
     rate of $0.4\times10^{-2}$pc$^{-3}$yr$^{-1}$  modulated 
     by 550~MV(top) and 850~MV(bottom)~\cite{Maki96,Yoshimura01}.}
    \label{fig:pbflux}
  \end{center}
\end{figure}

We obtained the antiproton flux at the TOA in the kinetic energy range 
0.10--4.20~GeV as shown in Fig.~\ref{fig:pbflux} and tabulated 
in Table~\ref{tab:pbflux}. 
The lowest energy was determined by the detector cutoff energy, 
and the highest energy was determined by the antiproton 
threshold energy of the aerogel \v{C}erenkov counter. 
The overall uncertainties including statistical and systematic errors are 
--34.1~\%/+41.8~\% at 0.16~GeV and $\pm$10.9~\% at 3.7~GeV 
with the given energy bin width. 
The statistical errors are dominant over the systematic errors below 1.4~GeV. 
Owing to the long duration observation the statistical errors 
were improved from the previous measurements with BESS, 
carried out on conventional 1 or 2 day balloon flights. 

Fig.~\ref{fig:pbflux} also shows the results from previous BESS flights 
around solar minimum (95+97)~\cite{Matsunaga98,Orito00} 
and maximum (2000)~\cite{Asaoka02}, which are compared with four 
theoretical calculations. 
The solid curves are calculations of secondary antiproton spectra 
with the Standard Leaky Box (SLB) model modulated with a steady state 
drift model~\cite{Bieber99}, in which the modulation is characterized by a 
tilt angle of the heliospheric current sheet and the Sun's magnetic 
polarity (denoted as +/-). The dashed curves are calculations 
with the Diffusion plus Convection (DC) model modulated with a drift 
model~\cite{Moskalenko02,Moskalenko07}. 
Three tilt angles, 10$^\circ(+)$, 70$^\circ(-)$, and 30$^\circ(-)$ 
roughly correspond to the measurements with BESS(95+97), 
BESS(2000) and BESS-Polar(2004), respectively~\cite{Zhao95,WSO}. 
The dotted curves are calculations with the DC model~\cite{Moskalenko02} 
modulated with a standard spherically symmetric approach~\cite{Fisk71}, 
in which the modulation is characterized by a single parameter ($\phi$) 
irrespective of the Sun's polarity. For each measurement, $\phi$ was 
obtained by fitting that proton spectrum measured by BESS, assuming the 
interstellar spectrum in Ref.~\cite{Shikaze07}. Three values of $\phi$, 
550~MV, 1400~MV, and 850~MV correspond to the measurements 
with BESS(95+97), BESS(2000) and BESS-Polar(2004), respectively. 
The dash-dot curves are calculations of antiproton spectra 
from evaporation of Primordial Black Holes (PBH) with an explosion 
rate of $0.4\times10^{-2}$pc$^{-3}$yr$^{-1}$~\cite{Maki96,Yoshimura01} 
modulated by the same spherically symmetric approach 
with $\phi$ of 550~MV and 850~MV. 
The expected signal from the PBH is 
modulated more than the secondary antiproton spectrum because of its 
spectral shape having the peak in the low energy region. 
While the BESS(95+97) data were suggestive of an excess flux 
at energies below 400 MeV, we do not find further evidence in the 
new BESS-Polar data presented here, even with the extended energy range. 

\begin{figure}
  \begin{center}
    \includegraphics[width=12cm]{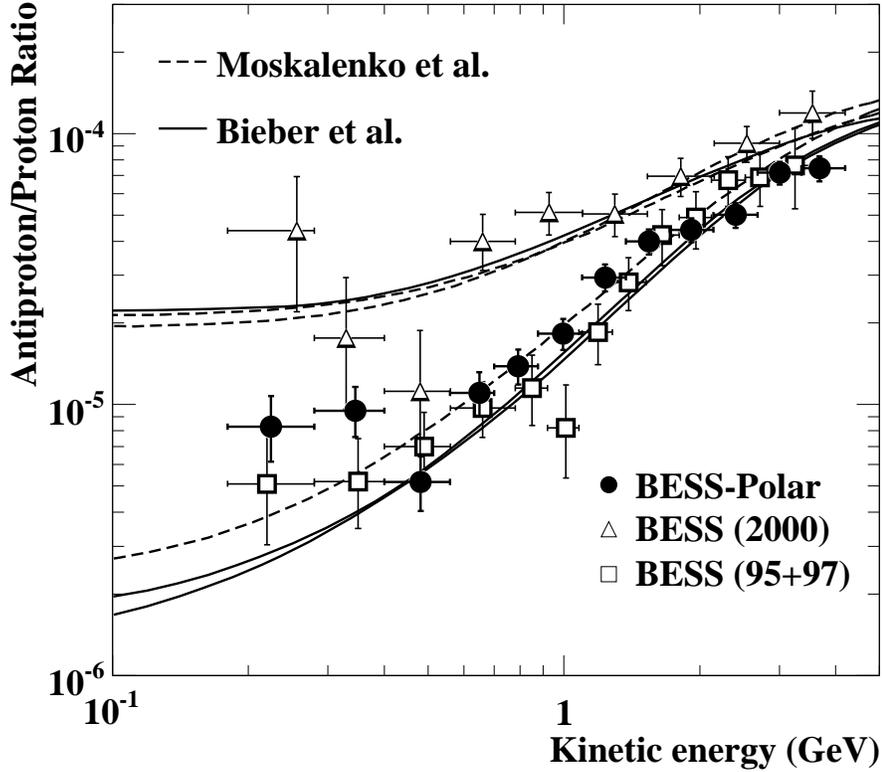}
    \caption
    {\={p}/p ratio obtained with the first 
     BESS-Polar flight together with results from previous BESS flights 
     around solar minimum (95+97)~\cite{Matsunaga98,Orito00}
     and maximum (2000)~\cite{Asaoka02}. 
     The solid curves are calculations with the Standard Leaky Box (SLB) model 
     modulated with a steady state drift model~\cite{Bieber99} by solar tilt 
     angles and magnetic polarities of (from bottom to top, the first two 
     are very close) 10$^\circ(+)$, 10$^\circ(-)$, and 70$^\circ(-)$. The 
     dashed curves are calculations with the Diffusion plus Convection (DC) 
     model~\cite{Moskalenko02} modulated by (from bottom to top, the last 
     two are very close) 10$^\circ(+)$, 30$^\circ(-)$, and 70$^\circ(-)$.}
    \label{fig:pbpratio}
  \end{center}
\end{figure}

\begin{figure}
  \begin{center}
    \includegraphics[width=12cm]{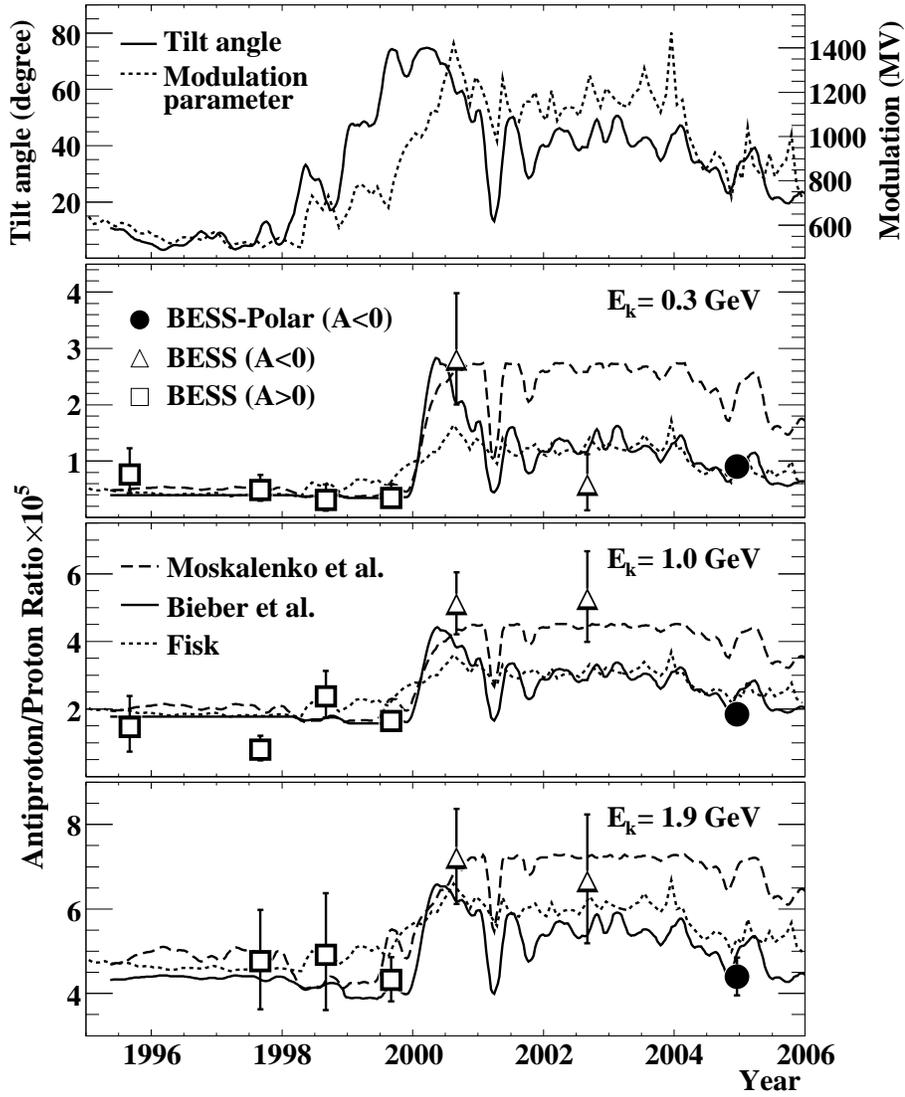}
    \caption
    {The top panel shows time variations of the tilt angle~\cite{Zhao95,WSO}  
     shown by a solid curve 
     and modulation parameter for the spherically symmetric 
     model~\cite{Fisk71} shown by a dotted curve. The modulation parameter 
     values continuous over time were estimated by using its linear 
     relation with Climax neutron monitor data~\cite{Climax}. 
     The linear relation was established using the modulation parameter 
     for each BESS flight obtained by fitting the BESS proton spectrum. 
     The other three panels show time variations of the \={p}/p ratio 
     at 0.3~GeV (2nd), 1.0~GeV (3rd), and 1.9~GeV (bottom).
     The data of \={p}/p ratio are compared with time variations predicted 
     by two drift models shown by solid curves from 
     Bieber et al.~\cite{Bieber99} and dashed curves from 
     Moskalenko et al.~\cite{Moskalenko02}, and with the 
     spherically symmetric moduation by Fisk~\cite{Fisk71} shown by 
     dotted curves.}
    \label{fig:timevar}
  \end{center}
\end{figure}

The \={p}/p ratio can provide a useful probe 
to study solar modulation and its charge-sign dependence. 
The BESS collaboration has tracked this ratio through most of a solar cycle. 
Fig.~\ref{fig:pbpratio} shows the \={p}/p ratio obtained with the 
BESS-Polar measurement together with the results from previous BESS flights 
around solar minimum (95+97)~\cite{Matsunaga98,Orito00} 
and maximum (2000)~\cite{Asaoka02}, which are compared with the same three 
calculations of secondary antiproton spectra as shown in Fig.~\ref{fig:pbflux}.
Fig.~\ref{fig:timevar} shows time variations of \={p}/p ratio at 
three different energies, 0.3~GeV, 1.0~GeV, and 1.9~GeV, each 
compared with the same calculations as shown in Fig.~\ref{fig:pbpratio}. 

Since the \={p}/p ratios obtained using drift models are given as a 
function of tilt angle, they were converted into those as a function of 
time by taking the tilt angle as the mean position of the monthly variation 
of the maximum latitudinal extent of the current sheet~\cite{Zhao95,WSO}.  
The time variation of \={p}/p ratios by the spherically symmetric approach 
were estimated by using a linear relation between $\phi_{BESS}$ and $N_{CL}$, 
where $\phi_{BESS}$ was obtained by fitting each proton spectrum 
measured with 6 BESS flights~\cite{Shikaze07} 
and $N_{CL}$ is the monthly averaged count rates of the Climax neutron
monitor~\cite{Climax}\footnote{U.S. National Science 
Foundation Grant ATM-0339527}.

Figs.~\ref{fig:pbpratio} and \ref{fig:timevar} show that drift models 
and the symmetric model reproduce equally well the stable \={p}/p ratio 
during the positive phase. The sudden increase of the ratio observed 
by BESS measurements after the positive-to-negative solar field reversal 
is better reproduced by the drift models at energies below 1~GeV. 
On the other hand, during the negative phase, where the \={p}/p ratio 
depends on the tilt angles more than during the positive phase, the 
two drift model calculations~\cite{Moskalenko02,Bieber99} significantly 
differ from each other. We found that the difference mainly comes 
from the difference in the modulation of protons, which in the negative 
phase should come from a ``horizontal'' direction in heliosphere, i.e. 
along the current sheet. This would imply 
that the model of particles behavior along the current sheet is not 
yet established well. Our data prefer calculations by Bieber et 
al.~\cite{Bieber99} but those with the spherical model are also preferred. 
In addition, Fig.~\ref{fig:pbflux} shows that 
the spherical model reproduces antiproton spectra of all the three 
BESS measurements better than drift models. 
This implies that the charge-sign dependence in the modulation 
is not so significant during the negative phase. The spherical model 
is still the best one to reproduce measurements.

\section{Conclusion}
\label{sec:conclusion}

The first BESS-Polar experiment was carried out in Antarctica in December 
2004, a transient period before the solar minimum in 2007. Using 
a new spectrometer with reduced material thickness and a long duration 
balloon flight near the Earth's south magnetic pole, 
the lowest energy limit of the antiproton flux was extended 
down to 0.1 GeV, and the statistics were improved compared 
with the previous BESS experiments. The series of BESS 
measurements have enabled a crucial test of models of the 
solar modulation. Drift models can reproduce the drastic behavior of the 
\={p}/p ratio around the solar field reversal, but for the negative phase 
the spherical model is still best able to reproduce most of the 
measurements. BESS data should motivate further development of the drift 
models with more realistic parameters and time dependence. For a more 
extensive future search for cosmic-ray antiprotons of primordial origin 
the BESS-Polar result provides an important baseline measurement of the 
secondary antiproton spectrum, which will be compared with a spectrum 
measured in December 2007 through January 2008, by the second Antarctic 
flight, during the solar minimum period. 

\begin{ack}
The authors thank NASA Headquarters for continuous 
encouragement in this U.S.-Japan cooperative project. Sincere thanks are 
expressed to the NASA Balloon Programs Office at GSFC/WFF and to the 
NASA Columbia Scientific Balloon Facility for their 
experienced support. They also thank ISAS/JAXA and KEK for their continuous 
support and encouragement. Special thanks go to the National Science 
Foundation (NSF), USA, and Raytheon Polar Service Company for their 
professional support in the USA and in Antarctica. 
The authors would thank the BESS-Polar II collaborators M.~Hasegawa, 
A.~Horikoshi, K.~Sakai, and N.~Thakur for their contribution to the 
instrument performance evaluation and further cooperation. The BESS-Polar 
experiment is being carried out as a Japan-U.S. collaboration. 
It is supported by MEXT grants 
(KAKENHI-13001004, 15340077, and 181040006) in Japan, 
and by NASA grants in the USA.
\end{ack}

\clearpage

\begin{table}[h]
  \renewcommand{\arraystretch}{1.4}
  \caption{Antiproton flux and \={p}/p ratio 
    at the top of atmosphere
    with statistical (first) and systematic (second) errors. 
    $N_{\bar{\rm p}}$ and $N_{BG}$ are the number of observed 
    antiprotons and estimated background events, respectively.}
  \vspace{0.5cm}
  \label{tab:pbflux}
  \begin{center}
    \begin{tabular}{cccccc} \hline \hline
      \multicolumn{2}{c}{
	\begin{tabular}{@{}cc@{}}       
	  \multicolumn{2}{c}{Kinetic energy (GeV)}\\
	  \hspace{0.3cm}range\hspace{0.3cm} & mean
	\end{tabular}
      } & 
      \begin{tabular}{@{}c@{}} $N_{\bar{\rm p}}$ \end{tabular} &
      \begin{tabular}{@{}c@{}} $N_{BG}$ \end{tabular} &
      \begin{tabular}{@{}c@{}}
	${\bar{\rm p}}$ flux\\
	(m$^{-2}$sr$^{-1}$s$^{-1}$GeV$^{-1}$)
      \end{tabular} &
      \begin{tabular}{@{}c@{}} \={p}/p ratio \end{tabular}\\
      \hline
0.10--0.18 & 0.16 &   8 &  0.0 & 4.37$^{+1.81+0.25}_{-1.47-0.25}\times 10^{-3}$ & \\
0.18--0.28 & 0.22 &  15 &  0.0 & 4.53$^{+1.31+0.27}_{-1.11-0.27}\times 10^{-3}$ & 8.28$^{+2.03+0.64}_{-2.39-0.64}\times 10^{-6}$\\
0.28--0.40 & 0.34 &  24 &  0.0 & 6.67$^{+1.48+0.27}_{-1.30-0.27}\times 10^{-3}$ & 9.49$^{+1.85+0.40}_{-2.10-0.40}\times 10^{-6}$\\
0.40--0.56 & 0.48 &  23 &  0.1 & 3.94$^{+0.91+0.30}_{-0.80-0.30}\times 10^{-3}$ & 5.18$^{+1.06+0.38}_{-1.20-0.38}\times 10^{-6}$\\
0.56--0.70 & 0.65 &  37 &  0.2 & 8.27$^{+1.53+0.47}_{-1.39-0.47}\times 10^{-3}$ & 1.10$^{+0.18+0.06}_{-0.20-0.06}\times 10^{-5}$\\
0.70--0.88 & 0.79 &  55 &  0.6 & 9.82$^{+1.41+0.59}_{-1.29-0.59}\times 10^{-3}$ & 1.38$^{+0.18+0.08}_{-0.20-0.08}\times 10^{-5}$\\
0.88--1.10 & 1.00 &  84 &  3.3 & 1.17$^{+0.14+0.07}_{-0.13-0.07}\times 10^{-2}$ & 1.83$^{+0.21+0.11}_{-0.22-0.11}\times 10^{-5}$\\
1.10--1.37 & 1.23 & 143 & 10.8 & 1.64$^{+0.15+0.12}_{-0.14-0.12}\times 10^{-2}$ & 2.94$^{+0.26+0.20}_{-0.27-0.20}\times 10^{-5}$\\
1.37--1.72 & 1.54 & 198 & 18.2 & 1.84$^{+0.15+0.13}_{-0.14-0.13}\times 10^{-2}$ & 4.00$^{+0.31+0.28}_{-0.33-0.28}\times 10^{-5}$\\
1.72--2.15 & 1.92 & 220 & 18.7 & 1.62$^{+0.11+0.13}_{-0.11-0.13}\times 10^{-2}$ & 4.42$^{+0.30+0.33}_{-0.30-0.33}\times 10^{-5}$\\
2.15--2.68 & 2.40 & 233 & 15.1 & 1.40$^{+0.09+0.12}_{-0.09-0.12}\times 10^{-2}$ & 5.01$^{+0.33+0.41}_{-0.33-0.41}\times 10^{-5}$\\
2.68--3.36 & 3.01 & 276 & 11.9 & 1.47$^{+0.09+0.11}_{-0.09-0.11}\times 10^{-2}$ & 7.17$^{+0.43+0.54}_{-0.43-0.54}\times 10^{-5}$\\
3.36--4.20 & 3.68 & 204 &  8.1 & 1.10$^{+0.08+0.09}_{-0.08-0.09}\times 10^{-2}$ & 7.46$^{+0.52+0.60}_{-0.52-0.60}\times 10^{-5}$\\
      \hline
      \hline
    \end{tabular}
  \end{center}
\end{table}

\clearpage

\clearpage

\clearpage

\clearpage

\clearpage

\end{document}